\documentclass[conference]{IEEEtran}
\usepackage{paralist}
\usepackage{tabularx}
\usepackage{booktabs}
\usepackage{graphicx}
\usepackage{adjustbox}
\usepackage{graphicx}
\usepackage{dblfloatfix}
\usepackage{xcolor}
\usepackage{url}
\usepackage[linesnumbered,ruled,vlined]{algorithm2e}

\usepackage{fancyhdr}
\usepackage{kantlipsum}
\fancypagestyle{plain}{
\fancyhf{}
\fancyhead[C]{Conference on \LaTeX} 

}
\usepackage{eso-pic}
\begin{document}
\AddToShipoutPictureBG*{
\AtPageUpperLeft{
\setlength\unitlength{1in}
\hspace*{\dimexpr0.5\paperwidth\relax}
\makebox(0,-0.75)[c]{\textbf{2022 IEEE/ACM International Conference on Advances in Social
Networks Analysis and Mining (ASONAM)}}}}

\title{Improving Code Review with GitHub Issue Tracking}

\author{
\IEEEauthorblockN
	{
			Abduljaleel Al-Rubaye
	}
	\IEEEauthorblockA
	{
		Department of Computer Science\\
		University of Central Florida\\
		Orlando, FL USA\\
		Email: aalrubaye@knights.ucf.edu\\
	}
\and
\IEEEauthorblockN{Gita Sukthankar}
\IEEEauthorblockA{\textit{Department of Computer Science} \\
\textit{University of Central Florida}\\
Orlando, FL \\
gitars@eecs.ucf.edu}
}
\maketitle
\IEEEoverridecommandlockouts
\IEEEpubid{\parbox{\columnwidth}{\vspace{8pt}
\makebox[\columnwidth][t]{IEEE/ACM ASONAM 2022, November 10-13, 2022}
\makebox[\columnwidth][t]{\url{http://dx.doi.org/10.1145/XXXXXXX.XXXXXXX}}
\makebox[\columnwidth][t]{978-1-6654-5661-6/22/\$31.00~\copyright\space2022 IEEE} \hfill}
\hspace{\columnsep}\makebox[\columnwidth]{}}
\IEEEpubidadjcol
\begin{abstract}
Software quality is an important problem for technology companies, since it substantially impacts the efficiency, usefulness, and maintainability of the final product; hence, code review is a must-do activity for software developers. During the code review process, senior engineers monitor other developers' work to spot possible problems and enforce coding standards.  One of the most widely used open-source software platforms, GitHub, attracts millions of developers who use it to store their projects. This study aims to analyze code quality on GitHub from the standpoint of code reviews. We examined the code review process using GitHub's \textit{Issues Tracker}, which allows team members to evaluate, discuss, and share their opinions on the proposed code before it is approved. Based on our analysis, we present a novel approach for improving the code review process by promoting \textit{regularity} and community involvement.
\end{abstract}
\begin{IEEEkeywords}
social coding platforms, GitHub, code review,
issue tracking
\end{IEEEkeywords}


\section{Introduction}
Quality is the degree to which a software implementation fulfills specifications and customer expectations. However, it may be argued that there is no uniform definition of code quality since developers have various ideas about what constitutes excellent code. When it comes to analyzing software and quantifying its quality, there are many different viewpoints to consider \cite{spinellis2006code}.  Common code desiderata include: extensibility, maintainability, readability, documentation, testability, efficiency, reliability, portability, and reusability. But almost everyone believes that code quality is a crucial concept, regardless of how we define it. 


\textit{Code review} is one of the best approaches for improving overall code quality. The quality measures mentioned above may not be feasible if the code is not adequately reviewed. GitHub's \textit{Issue tracking} system provides a process to manage the code base bugs, general project tasks, and action items collaboratively. 

\begin{figure}[htbp]
\centerline{\includegraphics[width=\columnwidth]{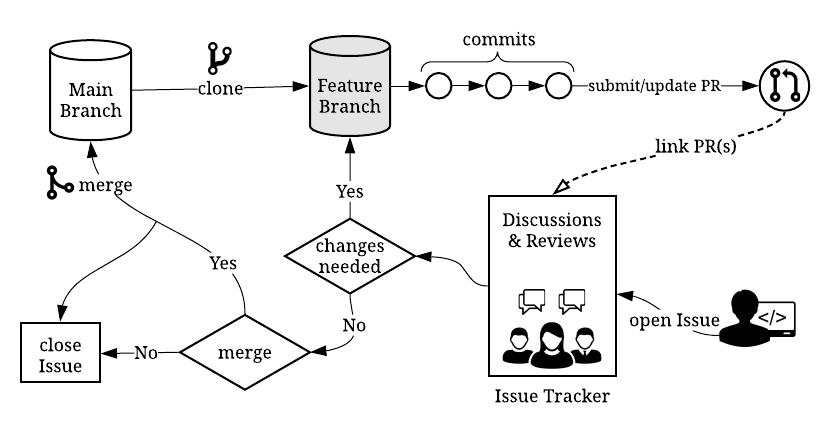}}
\caption{Issue tracking on GitHub.}
\label{issue_mechanism}
\end{figure}

Any GitHub user may start an Issue and moderate it. Discussion forums allow developers to track work, ask and answer questions, share expertise, and disseminate new ideas. Users can attach one or more pull requests to Issues for the team to evaluate before approving the new proposed modifications to the main code branch.  Figure~\ref{issue_mechanism} shows the overall flow of the GitHub Issue Tracker system.  Issues provide more than just a mechanism to report defects, since they allow others to participate and improve the work's quality. More significantly, the Issue tracking method helps participants comprehend the project's overall goal and disseminate new ideas for improving the code. 

This study aims to answer the following research questions using GitHub Issues:
\begin{compactitem}
    \item \textbf{RQ1}: How can we improve the regularity of the code review process?
    \item \textbf{RQ2}: What is the relationship between Issue frequency and community interaction?
    \item \textbf{RQ3}: How involved are the experienced reviewers in Issues?
    \item \textbf{RQ4}: Do more experienced GitHub users receive fewer comments?
\end{compactitem}

\section{Related Work}


Code quality is critical in the software development process, and no one can deny its significance. It is one of the most crucial stages in the ``done-ness'' of software. Ray et al. investigated the impact of programming languages on code quality in GitHub. After experimenting with various techniques and regression methods on bug problems reported on major GitHub projects, they discovered a direct correlation between code quality and the programming language used by developers~\cite{ray2014large}.

Several studies have considered adding features to social coding platforms to improve code quality.  Yu et al. examined the usage of GitHub's pull request mechanism to reduce potential bugs. They proposed a new comment-based approach to investigate developers' social communication before merging the code into the main code-base. When they combined their new proposed approach and GitHub's existing mechanisms, code quality improved~\cite{yu2016reviewer}. In another study, Yu et al. investigated the impact of user rules on code quality and how users' popularity can be related to the issues they introduce~\cite{lu2016does}.  The relationship between code review coverage and reviewers' engagement in software quality was examined by McIntosh et al. They looked at both current techniques and formal code reviews to demonstrate that both procedures are strongly linked to code quality~\cite{mcintosh2014impact}.  Our paper focuses exclusively on the effect of Issue tracking on the code review process and proposes a new mechanism for improving the existing system.



\section{Method}
For this project, we collected a large dataset of GitHub repositories and their associated events including Issues, Comments, and Commits\footnote{Available at \url{http://ial.eecs.ucf.edu/TeamComms/}}. To obtain our dataset, we used GHTorrent~\cite{gousios2013ghtorent}, an online archive of public repositories that collects all public data from GitHub on a daily basis.  We collected three categories of repositories:


\begin{compactenum}
\item \textbf{Random Repositories}:  We retrieved about 4,000 public repositories at random between March and June 2020.
\item \textbf{Popular Repositories}: The majority of the popular GitHub repositories are operated by large projects, organizations, or tech firms. These teams adhere to business criteria that maintain a high bar for code quality.  Thus, they may be a useful benchmark to compare to other GitHub repositories.  To identify a diverse collection of popular repositories, we first looked at The State of the Octoverse's list of the top five programming languages utilized on GitHub \cite{github_2020}. The top 20 repositories for each language were then picked using GitHub's Trending page, and the data was retrieved using the GitHub API \cite{github_api}.
\item \textbf{ROS-Related Repositories}: ROS, or Robot Operating System, is an open-source platform for robotics application development~\cite{quigley2009ros}. It provides a stable environment for coding complicated inter-component calculations. When creating robotics-related software, developers must maintain a high degree of quality, which led us to investigate GitHub repositories of this type and create a new benchmark to assess code quality.  \textit{GitHub topics}~\cite{github_topics} were used to find such repositories. They are subject-based labels that allow users to browse GitHub repositories by various categories. Using the GitHub API, we retrieved 3,000 public repositories with the term \textit{``ROS''} in at least one of their topics.
\end{compactenum}




Table~\ref{tab:statistics} shows statistics from our three different repository categories.  In comparison to the other two repo categories, the popular repositories have more Issues open, as seen in Table \ref{tab:statistics}. This is unsurprising and may be linked to the fact that large projects or developer teams have more \textit{contributors} than smaller projects or teams.


\begin{table*}[]
\begin{center}
	\caption{General statistics of the collected repositories.}
	\label{tab:statistics}
	\begin{tabular}{lrrr}
		\toprule
		General Statistics (Avg)&\textbf{Random Repos} & \textbf{ROS Repos} & \textbf{Popular Repos} \\
		\hline
		Issues per repository & $40$ & $130$ & $2155$ \\
		Closed Issues & $89\%$ & $88\%$ & $89\%$ \\
		Comments per closed Issue & $1.19$ & $1.68$ & $3.15$ \\
		Comments per non-closed Issue & $1.36$ & $1.95$ & $3.42$ \\
		Comments Sentiment Score $[-1,1]$ & $0.073$ & $0.085$ & $0.083$ \\
		Reviewers per Issues & $1.31$ & $1.42$ & $2.24$ \\
		Repository contributors & $5.5$ & $10$ & $146$ \\
        Repository owner followers & $81$ & $211$ & $771$\\
		Issue opener followers & $51$ & $82$ & $213$ \\
		Issue closer followers & $89$ & $185$ & $785$ \\
		Added / Removed lines per Issue & $1690$ / $670$ & $1300 / 650$ &$1330 / 432$\\
		Days prior to the first Issue & $128$ & $102$ & $71$ \\
		Hours to open an Issue & $450.94$ & $170.5$ & $25.2$\\
		Hours to close an Issue & $13.88$ & $15.61$ & $15.38$ \\
        Commits per repository & $216$ & $695$ & $10776$ \\
		Commits prior the initial Issue & $83$ & $102$ & $753$\\
		Commits between Issues & $5.42$ & $6.35$ & $8.48$ \\

		\bottomrule
	\end{tabular}
\end{center}
\end{table*}

Although large teams generate greater numbers of Issues, the code review process is still completed by a small number of \textit{reviewers}: the contributors who review and comment on code through Issues. On ROS-related and Random repositories, the average number of reviewers is slightly lower than that of large teams on Popular repositories. This might explain why, on average, all types of gathered repositories have approximately the same amount of \textit{comments} per issue.  The \textit{Sentiment Score} of an Issue comment indicates whether the comment is likely to be positive or negative. This number ranges from -1 to 1, with the greater the sentiment score, the more positive the comment's context. To calculate this number, we used TextBlob, a Python sentiment analysis tool~ \cite{loria2018textblob}.

Given that developers submit code updates through \textit{commits} before each Issue, it's worth noting that the number of \textit{added} and \textit{removed} lines of code in Random repositories is slightly greater than in ROS-related or Popular ones. To put it another way, developers on Random repositories are more likely to make a large number of code changes before committing them to the repository. When it comes to code review, however, Issues on Random repositories receive fewer comments than other types of repositories. When comparing the lines of new code to the number of comments through open Issues, this finding might be an indicator of lower quality work on Random repositories.

Developer activity is a good place to start examining Issues on GitHub repositories. As shown in Table \ref{tab:issues_correlations}, teams with a higher number of contributors are more likely to have more Issues than repositories with a lower number of contributors. Popular vs. less popular repositories exhibit a similar pattern: repositories with more Stargazers, Watchers, or Forkees are more likely to have more Issues opened.  In addition, the number of total reviewers on a repository is related to the number of Issues.  More Issues being opened may result in increased Contributor participation, suggesting improved collaboration and teamwork.



\begin{table*}[]
\begin{center}
	\caption{Correlation between the total Issue count and other features for all three types of repositories.}
	\label{tab:issues_correlations}
\begin{tabular}{lrl|rl|rl}

                                                    & \multicolumn{2}{c}{\textbf{Random Repos}} & \multicolumn{2}{c}{\textbf{ROS Repos}} & \multicolumn{2}{c}{\textbf{Popular Repos}} \\ 
                                                    & \textit{R-Val}& \textit{P-Val}& \textit{R-Val}& \textit{P-Val}& \textit{R-Val}& \textit{P-Val}\\ \bottomrule
\multicolumn{1}{l|}{Repo Age}                       & $0.217$       & $1.28^{-44}$  & $0.216$       & $1.89^{-33}$  & $0.406$       & $1.54^{-5}$   \\ 
\multicolumn{1}{l|}{Contributors}                   & $0.46$        & $8.75^{-213}$ & $0.812$       & $0$           & $0.765$       & $7.59^{-15}$  \\ 
\multicolumn{1}{l|}{Issue Comments}                 & $0.914$       & $0$           & $0.925$       & $0$           & $0.781$       & $4.93^{-21}$  \\ 
\multicolumn{1}{l|}{Commits}                        & $0.438$       & $8.43^{-191}$ & $0.728$       & $0$           & $0.451$       & $1.21^{-6}$   \\ 
\multicolumn{1}{l|}{Reviewers}                      & $0.194$       & $8.56^{-36}$  & $0.194$       & $3.41^{-27}$  & $0.254$       & $0.008$       \\ 
\multicolumn{1}{l|}{Commits before the first Issue} & $0.058$       & $0.00019$     & $0.166$       & $4.15^{-20}$  & $0.139$       & $0.154$       \\ 
\multicolumn{1}{l|}{Commits between Issues}         & $-0.019$      & $0.216$       & $-0.026$      & $0.142$       & $0.107$       & $0.271$       \\ 
\multicolumn{1}{l|}{Issue Opener Followers Count}   & $0.053$       & $0.0006$      & $0.045$       & $0.129$       & $0.142$       & $0.146$       \\ 
\multicolumn{1}{l|}{Repo Owner Followers Count}     & $0.077$       & $6.9^{-7}$    & $-0.013$      & $0.446$       & $-0.246$      & $0.010$       \\ 
\multicolumn{1}{l|}{Stars}                          & $0.387$       & $5.76^{-146}$ & $0.457$       & $7.79^{-156}$ & $0.311$       & $0.001$       \\ 
\multicolumn{1}{l|}{Forks}                          & $0.351$       & $1.19^{-118}$ & $0.409$       & $1.43^{-122}$ & $0.452$       & $1.13^{-6}$   \\ 
\multicolumn{1}{l|}{Watchers}                       & $0.391$       & $6.73^{-149}$ & $0.404$       & $4.18^{-119}$ & $0.328$       & $5.7^{-4}$    \\ 
\bottomrule
\end{tabular}
\end{center}
\end{table*}

\begin{figure}
\centerline{\includegraphics[width=0.8\columnwidth]{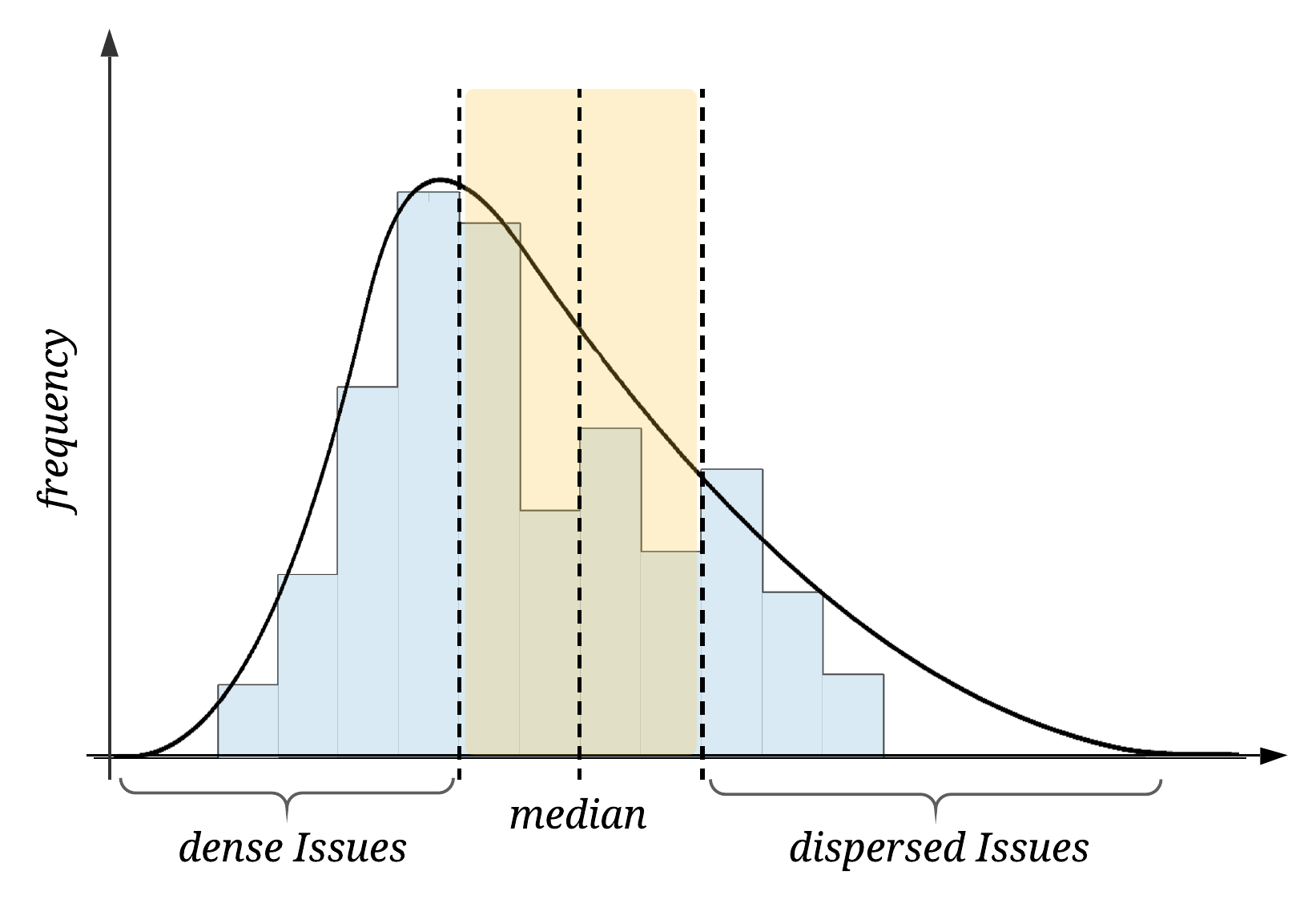}}
\caption{Time interval between open Issues}
\label{issues_distribution}
\end{figure}




Rather than examining the total number of Issues, we hypothesize that examining the time intervals between GitHub Issue opening events might provide a more accurate picture of how a team's activities affect code quality. We have divided the Issues into three categories: dense, regular, and dispersed, based on how close they are to the median of the time distance distribution.

Dense issues reflect rapid team activity through Issue Trackers. Contributors participate in activities and conversations to examine and evaluate new code quickly through various Issue events. A variety of factors might cause many issues to arise within a short period. This is, nevertheless, normal team behavior, particularly when a significant new feature is merged into the main branch via a series of pull requests. As a result, the team will have to devote extra time to evaluate the code quality before it can be accepted.

On the other hand, having numerous consecutive Issue events that are separated from one other suggests that there are fewer social interactions amongst the contributors of a GitHub repository over time. In such cases, contributors may not have completed a code implementation to discuss through Issue tracking. When a team conducts multiple code merges outside of the Issue tracking system, while the rate of dispersed Issues is high, it can result in a substantial gap in knowledge transfer, technical conversations, and thought exchange, all of which can influence the team's overall code quality.

Issues in the Regular distance range, on the other hand, suggest continuous code review behavior, in which contributors frequently interact through Issue Trackers. We refer to this team behavior as \textbf{\textit{Regular Code Reviews}}. Repositories with a greater rate of Regular Issues have a higher chance of performing regular quality checks, which helps to maintain and improve the code's high quality. According to \cite{code_review_survey}, code review has been ranked as the most effective way to achieve better quality.  Several annual studies on software quality looked at code reviews and attempted to determine which elements had the most impact on quality improvement. According to \cite{code_review_survey}, regular code review has been the most effective approach for improving quality; the next section describes our proposed mechanism for promoting regular code review.

\section{Results}
This section presents the results of our investigation into the Issue Tracking system and the code review process.
\subsection{Issue Regularity}

\smallskip
\smallskip
\textbf{RQ1: How can we improve the regularity of the code review process?}
\smallskip
\smallskip

\begin{figure}[htbp]
	\begin{center}
	\includegraphics[width=\columnwidth]{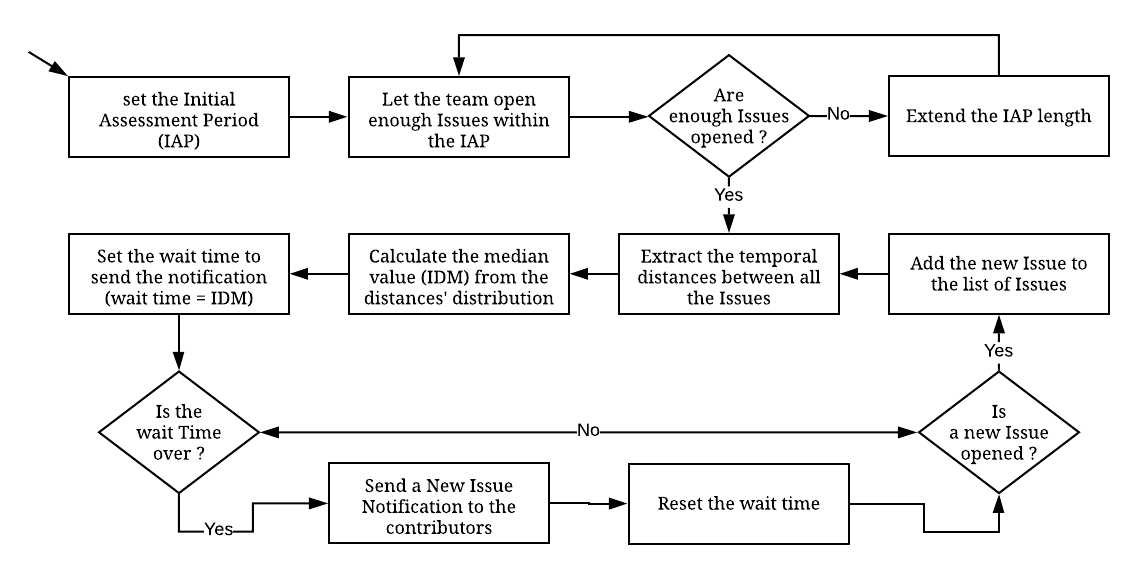}
	\caption{Proposed Issue Notifier (NIN) mechanism}
	\label{notification_diagram}
	\end{center}
\end{figure}

The code quality of GitHub projects is affected by a number of variables. As previously discussed, one of the most important approaches to enhance quality is to perform frequent code reviews, which may be conducted through the GitHub Issues Tracking system.  To do this, we study the repositories' timeline and investigate the distribution of temporal distances between Issues to create a reminder mechanism to attain \textit{Regularity}.  We propose the \textit{New Issue Notifier} (\textit{NIN}), a new mechanism that assesses contributors' behavior over time (Figure~\ref{notification_diagram}). To determine the timing of prompts, NIN leverages the median of the Issues temporal distances distribution. The suggested time to open an Issue is not static, as we continually recalculate it based on the median over time. We use this method to decide when to notify the team, reminding them that it may be time to open an Issue, which they may either accept or decline.

\begin{figure*}[htbp]
	\begin{center}
	\includegraphics[width=2.0\columnwidth]{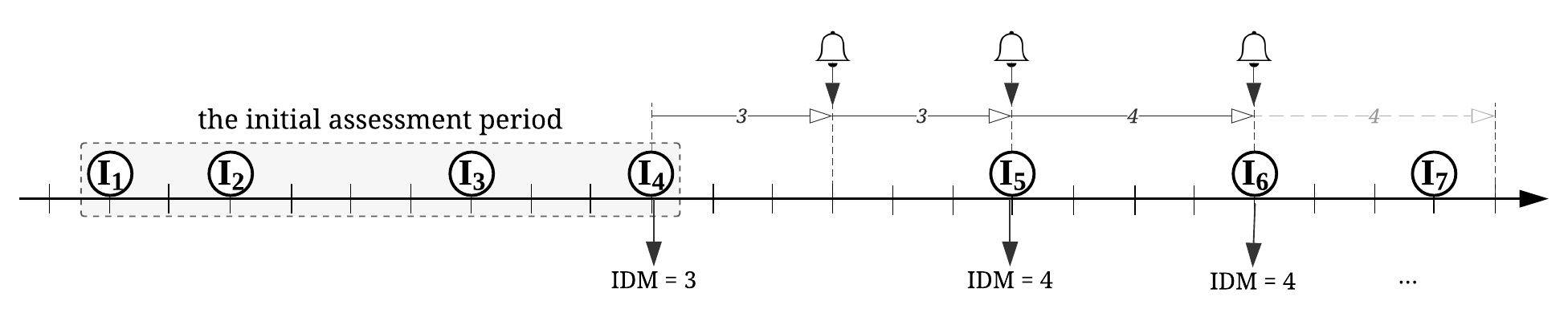}
		\caption{New Issue Notifier mechanism (NIN) applied to a GitHub repository timeline. IDM stands for Issue Distances Median.}
	\label{notification}
	\end{center}
\end{figure*}

The process starts by retrieving the history of the repositories' Issue Opening events, as illustrated in Figure \ref{notification}. To initialize the system, we need enough Issue-related data to begin evaluating the teams' behavior regarding opening Issues. The \textit{initial assessment period} is defined as the time frame during which we monitor the teams' activities in terms of opening new Issues for this purpose. The temporal distances between the Issues opened during the assessment time frame are extracted at the end of this period.  Afterward, we calculate the Issue Distances Median ($IDM$), which is the average time contributors spend opening new Issues throughout the development's life cycle.

To determine the best time for the team to open a new Issue, we utilize the IDM which is extracted from the team's activity pattern. The IDM is considered to be the waiting time to push a notification to the contributors if the team has not already opened an Issue. The notification suggests opening a new Issue; however, the team can act upon the notification and start a new Issue or ignore it and continue without opening any Issue. The notification will be scheduled to trigger repeatedly after the waiting time is over. If the team opens a new Issue (before or at the notification time), the IDM is calculated again, but this time for a bigger set of Issues, including the new one. The more Issue data we have, the more accurate the IDM becomes.
We developed a simulation of the New Issues Notifier, or \textit{NIN}, considering the following key factors:

\begin{compactitem}
\item The examination included all repositories from the three repository categories.
\item All repositories with an insufficient number of Issues were deemed noisy data and were thus eliminated before the simulator was run.
\item The simulator was fed a collection of actual ground truth Issue data from each repository for the initial assessment period as its first input.

\item Table \ref{tab:statistics} shows that contributors wait an average of 2 to 4 months before opening the first Issue across all three types of repositories. After that, the pace of opening Issues rises, and they tend to open Issues several times per month. As a result, the initial assessment period (IAP) has been set at six months from the day the repository was created in order to allow us to collect adequate data.

\item We defined three \textit{Acceptance Probabilities} (AP): $0.3$, $0.6$, and $0.9$. They reflect the likelihood that the team will accept the New Issue notification.
\end{compactitem}

The simulator was used to recreate three years of GitHub team activity. The simulator operates on three separate threads, each with its acceptance probability (IAP). Once the execution was completed, we retrieved the distribution of our repositories' Dense, Regular, and Dispersed Issues. As indicated in Table \ref{tab:percentage_table}, we can see an increase in the number of Issues that are opened on a more frequent basis.

 \begin{table}[]
 \begin{center}
 	\caption{The percentage of dense, regular, and dispersed Issues after running the simulator for each acceptance probability (AP).}
	\label{tab:percentage_table}
\begin{tabular}{l|l|ccc}
\bottomrule
 & AP & \multicolumn{1}{l}{Dense} & \multicolumn{1}{l}{Regular} & \multicolumn{1}{l}{Dispersed} \\ \bottomrule
 & $0.3$ & $9.60$\% & $79.63$\% & $10.77$\% \\
Popular Repos & $0.6$ & $9.78$\% & $81.13$\% & $9.09$\% \\
& $0.9$ & $10.52$\% & $81.10$\% & $8.38$\% \\ \hline
& $0.3$ & $16.15$\% & $66.54$\% & $17.31$\% \\
ROS Repos & $0.6$ & $14.26$\% & $70.90$\% & $14.83$\% \\
& $0.9$ & $13.39$\% & $74.47$\% & $12.14$\% \\ \hline
& $0.3$ & $12.12$\% & $69.45$\% & $18.43$\% \\
Random Repos & $0.6$ & $10.89$\% & $72.25$\% & $16.86$\% \\
& $0.9$ & $9.99$\% & $74.95$\% & $15.06$\% \\ \bottomrule
\end{tabular}
\end{center}
\end{table}

\begin{figure*}[htbp]
	\begin{center}
	\includegraphics[width=1.5\columnwidth]{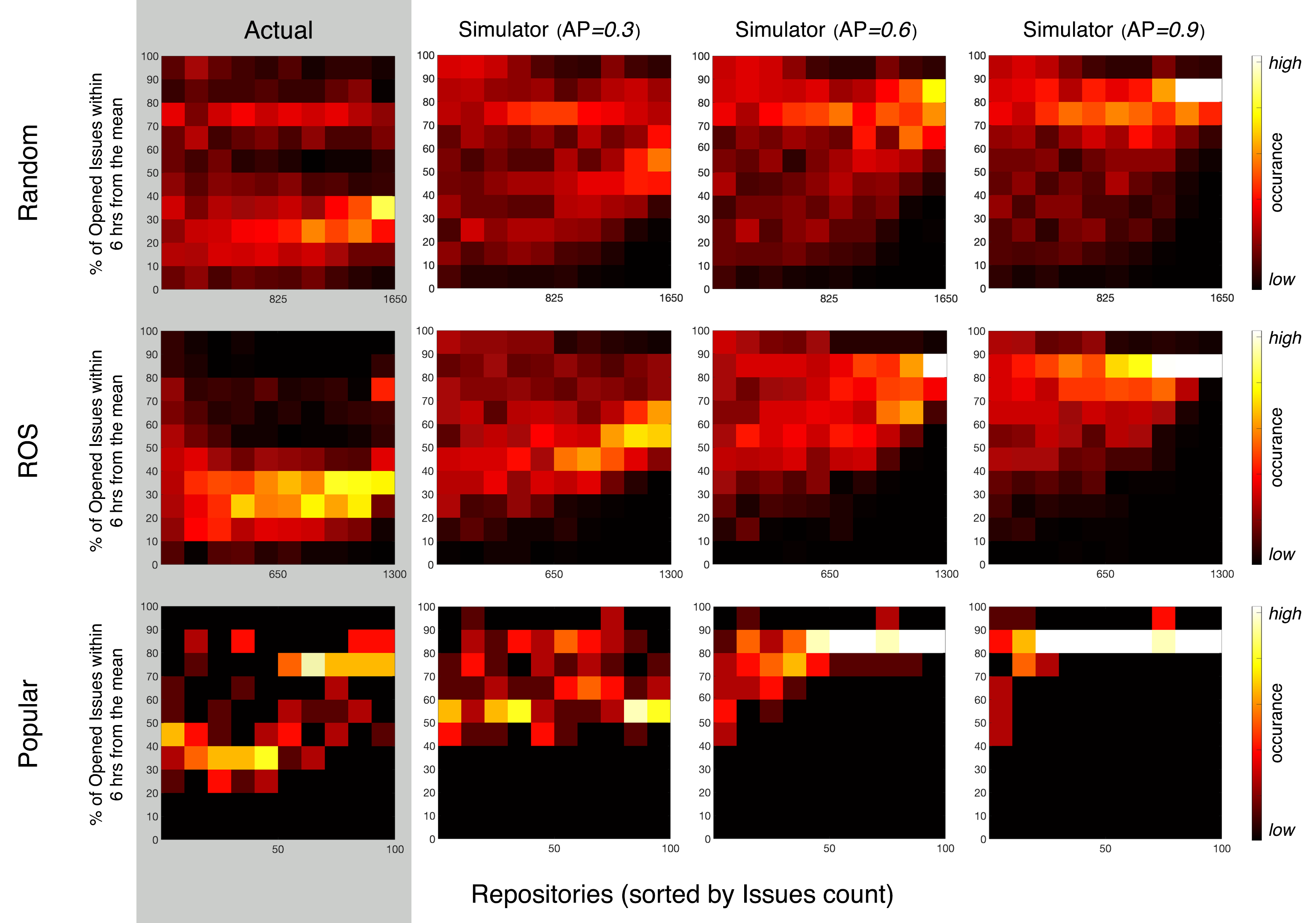}
	\caption{The percentage of the Issues that fall within the regular area [$median - \alpha, median + \alpha$], where $\alpha$ = 6 hours, for all three types of the repositories before and after running the simulator (for three different acceptance probabilities). }
	\label{heatMap}
	\end{center}
\end{figure*}

A higher acceptance ratio increases the numbers of Regular Issues. This change is slightly less on Popular repositories compared to ROS and Random repositories. Figure \ref{heatMap} shows a collection of heat maps depicting the percentage of Issues that are opened more regularly before and after the simulator was performed. The figure indicates that the Regular Issues rate rises over time as more Issues are opened, irrespective of the chance of notification acceptance.  The New Issue Notifier method, (NIN), is an approach that uses the Issue Tracking system to improve regularity. It nudges the team to collaborate more frequently to discuss the new implementation and review the submitted code before merging, which may improve code quality over time.

\subsection{Issues Community}
\smallskip
\smallskip
\textbf{RQ2: What is the relationship between Issue frequency and community interaction?}
\smallskip
\smallskip

Contributors open Issues on GitHub to allow the team to collaborate and exchange information. Issues may be seen as parts of a larger community within a repository from a broad perspective. As a result, when it comes to quality, we should take into account the larger community, where users contribute to a greater goal: a \textit{high-quality product}. We constructed a network of Issues and their reviewers to better comprehend users' engagement in repositories' Issues and explore their interactions. This network is a multi-layer network, with node connections on the first layer defining connectivity on the second layer. From the standpoint of code review, this network reflects the Issues Community of GitHub repositories. The network is formed as follows: 

\begin{itemize}
\item Node type 1 ($n_1$): represents repository $R$'s opened Issues $\{ I_1, I_2, ..., I_n \}$, where $n$ is the total number of the Issues.

\item Node type 2 ($n_2$): represents the users $\{r_1, r_2, ..., r_m\}$ who reviewed the proposed code via Issues in repository $R$, where $m$ is the total number of the reviewers.

\item Edge type 1 ($e_1$): a weighted link that connects $r_x$ ($n_2$ nodes) to $I_y$ ($n_1$ nodes), if $r_x$ has reviewed the code through $I_y$, where $1 \le x \le m$, and $1 \le y \le n$. The more comments $r_x$ has through $I_y$, the greater weight the edge ($r_x$ - $I_y$) gets. 

\item Edge type 2 ($e_2$): a weighted link that connects $I_i$ to $I_j$ ($n_1$ nodes) if $r_x$ has reviewed codes through both Issues $I_i$ and $I_j$. 
\end{itemize}

\begin{figure}[htbp]
	\begin{center}
	\includegraphics[width=0.9\columnwidth]{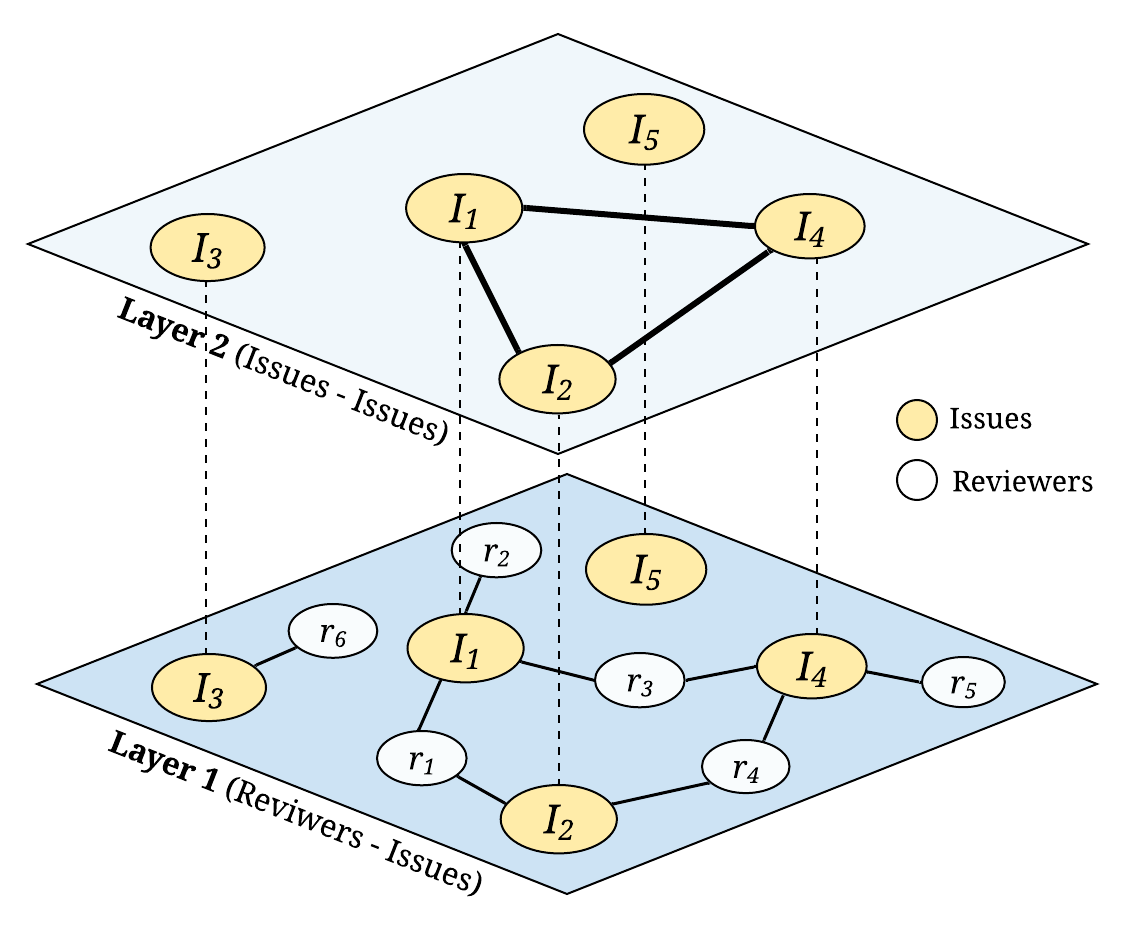}
	\caption{A sample Issue Community (IC) network is formed by two types of nodes that represent Issues, and reviewers. There are two layers: Layer 1 shows the edges of type $e_1$ that connect Reviewers to Issues, and Layer 2 shows the edges of type $e_2$ that connect Issue to Issue. }
	\label{network_Sample}
	\end{center}
\end{figure}

\begin{figure*}[htbp]
	\begin{center}
	\includegraphics[width=1.8\columnwidth]{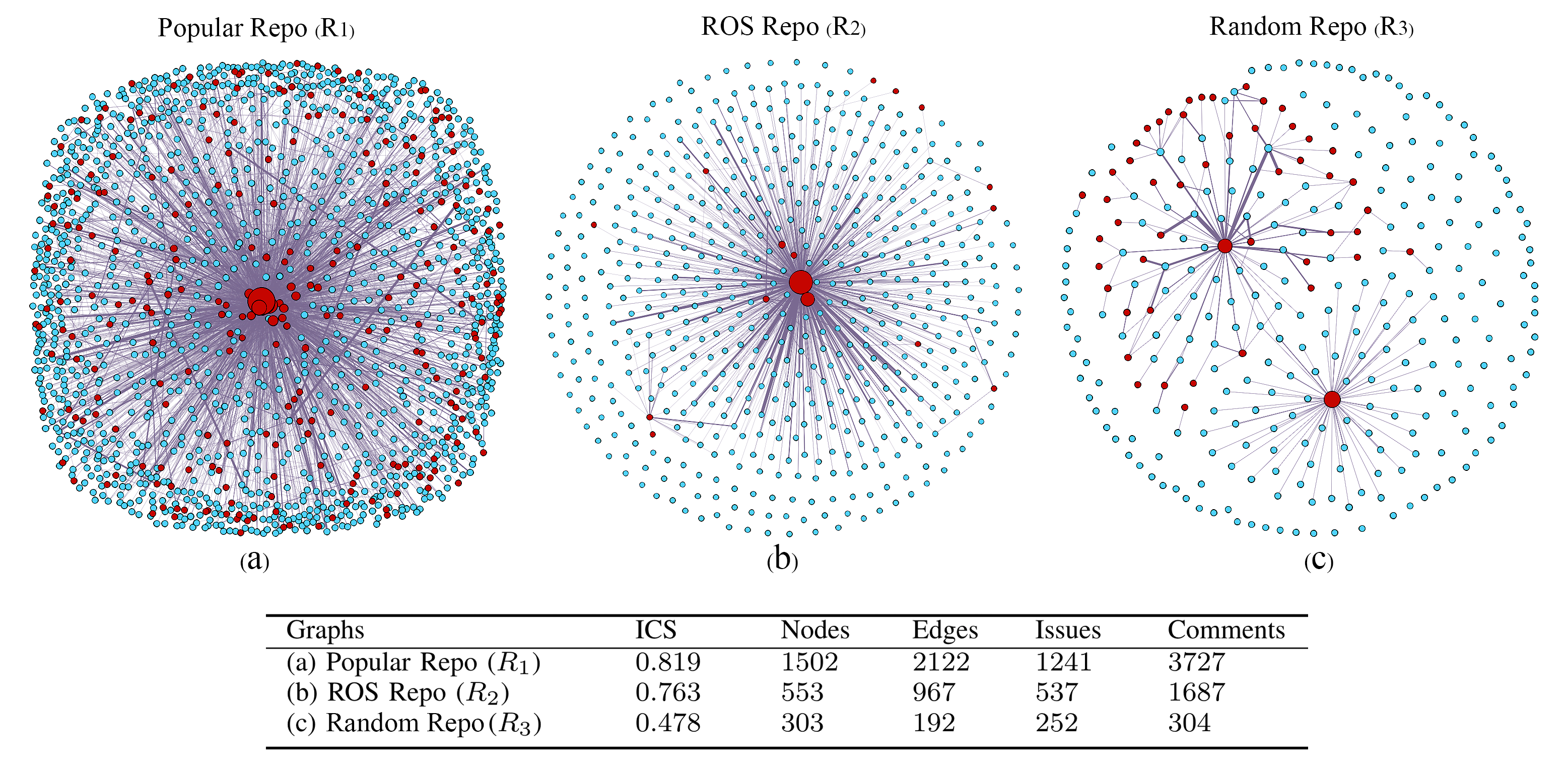}
	\caption{Subgraphs of three different Issue Communities, with just the $e_1$ edges shown (layer 1 connections Reviewers-Issues). The reviewers are represented by the red nodes, while the open Issues are represented by the blue nodes. The size of the node reflects its degree.}
	\label{IssueCommunityGraphs}
	\end{center}
\end{figure*}

Figure \ref{network_Sample} shows a sample Issue Community of repository $R$, with reviewer nodes $r 1$, $r 3$, and $r 4$ connected to several open Issues. In comparison to others, these nodes indicate contributors who can provide help across Issues. These reviewers are more likely to participate in conversations, share their opinions, and, in general, help their teammates evaluate their work and merge any code that meets high-quality standards. Figure \ref{IssueCommunityGraphs} depicts three separate Issue Communities along with their scores.  The graphs only show edges of type $E_1$ to make them more visually clear.

Reviewer collaboration behaviors vary depending on the repository and the project they are working on. Some repositories have proactive and experienced team members who are willing to participate in technical reviews and conversations for the team's benefit, while others have less active users. As a result, we defined the \textit{Issue Community Score} (\textit{ICS}) to assess each repository's Issues Community in terms of user involvement. The \textit{ICS} represents the flow of information, idea sharing, thought interchange, and the interconnectedness between Issues and Reviewers. We have calculated the ICS for each repository as follows: 
$$ICS_R = \frac{\left | IC_R(e_2) \right |}{\left |Issues_R - 1  \right |}$$
where the numerator is the total number of the edges of type ($e_2$) within the Issue Community of the Repository $R$. If all of the Issues nodes in a community are connected, it receives a perfect score. A higher score indicates the extent to which contributors' expertise and knowledge are shared through Issues. A lower \textit{ICS}, on the other hand, shows that there are distinct Issues that are not connected to other parts of the community.  Figure \ref{network_Sample} shows an example in which $I_3$ and $I_5$ are separated from other Issues. $I_5$ is reviewed by $r_6$ who is not involved in other issues, while $I_5$ is opened and closed with no reviews.

Table \ref{tab:correlactions_community_score} shows the correlation between \textit{ICS} and other features. In comparison to other characteristics like repository contributors count, stargazers, or the total number of Issues, we found that \textit{ICS} strongly correlates with the total number of Issues comments. This correlation appears to be stronger on popular repositories, suggesting that repository contributors use Issues to communicate, share ideas, and review codes. (See Table \ref{tab:statistics}).

Unfortunately, teams who open Issues on a regular basis may not always have higher code quality. From the standpoint of Issue tracking, greater code quality necessitates not only the opening of more Regular Issues, but also the participation of more contributors through Issues and collaboration as a single team. Investigating our data set, we have noticed that there are repositories that have many of their Issues opened and closed with only a couple or no reviews at all. User participation, involvement, and communication is as crucial as the concept of Issue Regularity.

\subsection{Expertise Coverage}
\textbf{RQ3: How involved are the experienced reviewers in Issues?}

When code is submitted through an Issue to be evaluated and discussed by other contributors, it is critical that the highly experienced team members participate in the process. With their seniority, expertise, ideas, and general high-level vision, experienced contributors guide the team.  The quality of the work will be improved if more of these users contribute directly to a larger number of Issues. A repository with highest number of Issues reviewed by experienced individuals is likely to have better code quality than one with the most of its Issues reviewed by less experienced team members. We studied the Expertise Issue Coverage on different types of repositories based on this assumption.

\begin{table}[ht]
\begin{center}
	\caption{Follower counts for users}
	\label{tab:code_reviewers_popularity}
	\scalebox{0.9}{
        \begin{tabular}{lccc}
		\toprule
		&Random Repos & ROS Repos & Popular Repos \\
		\hline
		Issue opener followers & 51 & 82 & 213 \\
		Issue closer followers & 89 & 185 & 785 \\
        Reviewer followers & 71 & 180 & 542 \\
		\bottomrule
	\end{tabular}}
\end{center}
\end{table}

When developers register a GitHub account and begin their coding adventure, their publicly available coding material usually attracts other users. In GitHub users can follow other developers and receive notifications when the followee publishes new content; users are likely to be attracted to more experienced developers who produce higher quality work. As GitHub users maintain a high level of competence, their popularity rises over time, and they acquire more experience.  As a result, the number of people who follow a user is roughly proportional to their seniority and experience level. Table \ref{tab:code_reviewers_popularity} shows the reviewers' followers by repository type, as well as followers for users who open and close Issues.

\begin{figure}[htbp]
	\begin{center}
	\includegraphics[width=0.8\columnwidth]{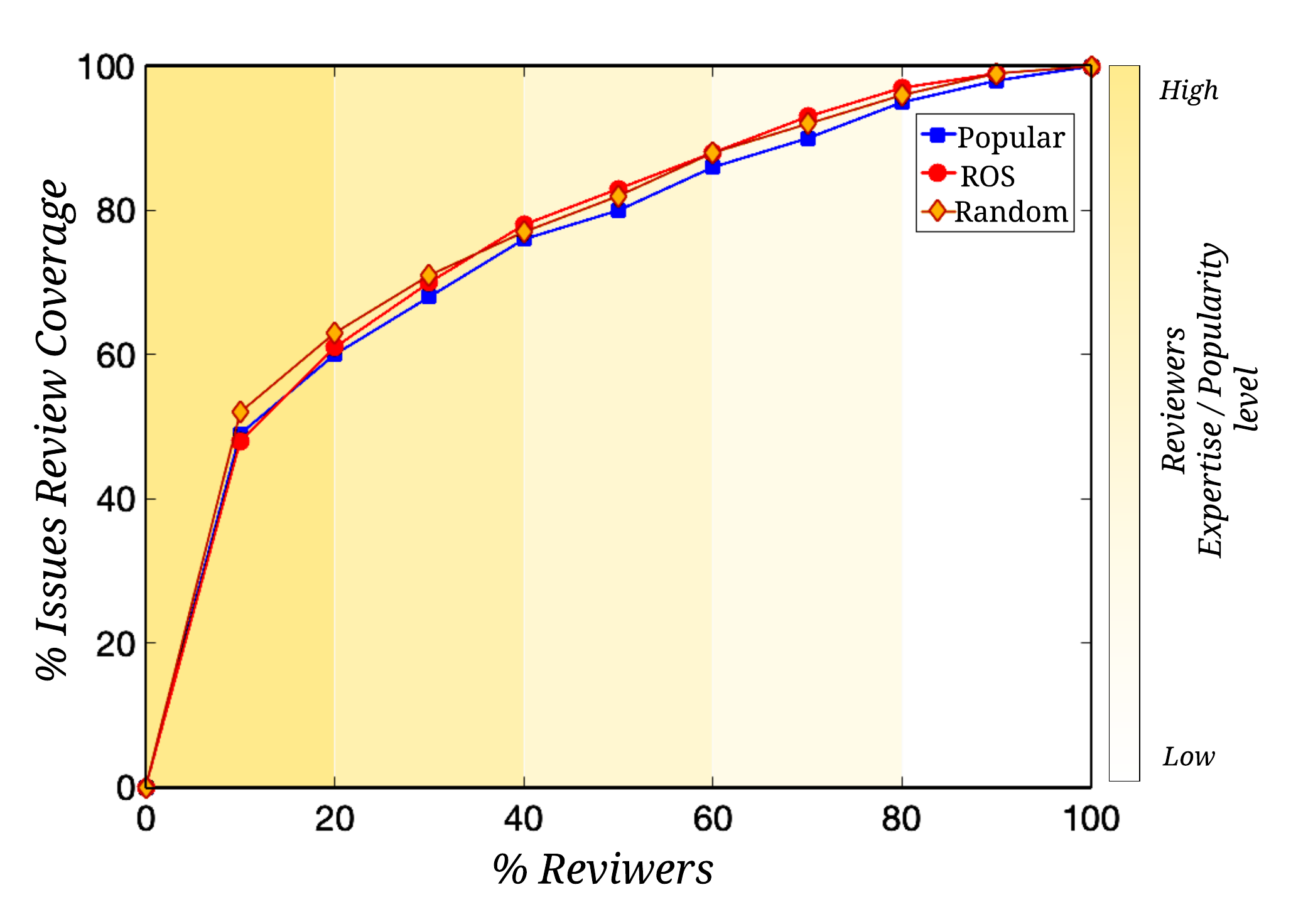}
	\caption{Issues coverage by Reviewers for all three types of repositories.  Reviewers are sorted from high to low by their popularity.}
	\label{expertise-coverage}
	\end{center}
\end{figure}

The Issue coverage by reviewers is depicted in Figure \ref{expertise-coverage}, with reviewers ordered from high experience to low experience. The graph demonstrates that all three types of repositories follow the same overall expertise coverage pattern, with the most popular and experienced users representing roughly 20\% of the total reviewers.  Table \ref{tab:expertise_coverage} shows that on average, 90\% of a repository's users' followers follow the most popular 20\% of the reviewers. Regardless of repository type, the 20\% most popular reviewers cover around 60\% of all Issues within their repositories to assess other team members' work and provide feedback and correction. Despite the fact that the data indicates a similar Issue coverage pattern for Popular, ROS, and Random repositories, the levels of popularity and expertise vary. As shown in Table \ref{tab:code_reviewers_popularity} on Popular repositories, contributors closing an Issue are far more popular and thus experienced than on ROS and Random repositories. As a result, regardless of the similar coverage pattern, the quality of the review and assistance on the Popular repositories might be significantly greater.  Nonetheless, the expertise coverage percentage shows that when it comes to code reviews, popular reviewers play an essential role in spreading their opinions, ideas, and experience through Issues, which assists other less experienced developers in learning how to improve their work quality.

\begin{table}[]
\begin{center}
	\caption{Popularity ratio and the percentage of Issues that 20\% of the most popular reviewers cover. On average, 20\% of the most popular cover more than half of the Issues.}
	\label{tab:expertise_coverage}
\begin{tabular}{lll}
\bottomrule
Repositories Type & Popularity Ratio & Issue Coverage \\ \hline
Popular Repos & 93\% & 59\% \\
ROS Repos& 90\% & 60\% \\
Random Repos& 87\% & 61\% \\ \bottomrule
\end{tabular}
\end{center}
\end{table}

\begin{table}[]
\begin{center}
	\caption{Correlation of repository properties: Issues, Contributors, Stars, Issue Comments and the Issue Community Score. }
	\label{tab:correlactions_community_score}
	\scalebox{0.9}{
\begin{tabular}{lrl|rl|rl}
& \multicolumn{2}{c}{\textbf{Random Repos}} & \multicolumn{2}{c}{\textbf{ROS Repos}} & \multicolumn{2}{c}{\textbf{Popular Repos}} \\
& \textit{R-Val}& \textit{P-Val}& \textit{R-Val}& \textit{P-Val}& \textit{R-Val}& \textit{P-Val}\\ \bottomrule
\multicolumn{1}{l|}{Contributors}& $0.146$&$0.061$& $0.226$& $0.001$ & $0.296$& $2.788^{-3}$\\ 
\multicolumn{1}{l|}{Stargazers}& $0.124$& $0.113$& $0.235$& $7.703^{-4}$& $0.167$& $0.095$  \\ 
\multicolumn{1}{l|}{Issues Count}& $0.219$& $0.004$ &$0.115$& $0.102$& $0.227$& $0.022$\\ 
\multicolumn{1}{l|}{Comments}& $0.359$& $2.33^{-7}$& $0.259$& $1.966^{-4}$& $0.427$& $9.17^{-6}$\\ 
\bottomrule
\end{tabular}}
\end{center}
\end{table}

\subsection{Developer Popularity}
\textbf{RQ4: Do more experienced GitHub users receive fewer comments?}

When a contributor opens an Issue and links a pull request to it, others can participate in reviewing, share their thoughts, and suggest updating the code if a revision is needed. Now imagine if a high-quality piece of code is linked to an Issue for review. This code may receive fewer comments from the reviewers compared to code that needs to be updated. Many factors can impact the quality of the code, but we assume that the seniority and the developer's experience can significantly impact his/her code's quality.

To address research question 4, we examine the relationship between Issues openers' popularity and the comments they receive through Issues, taking into account that a user's popularity is determined by the number of followers they have. Only Issues in which their openers had submitted their code for review were included in this investigation.


\begin{figure}[htbp]
	\begin{center}
	\includegraphics[width=1.0\columnwidth]{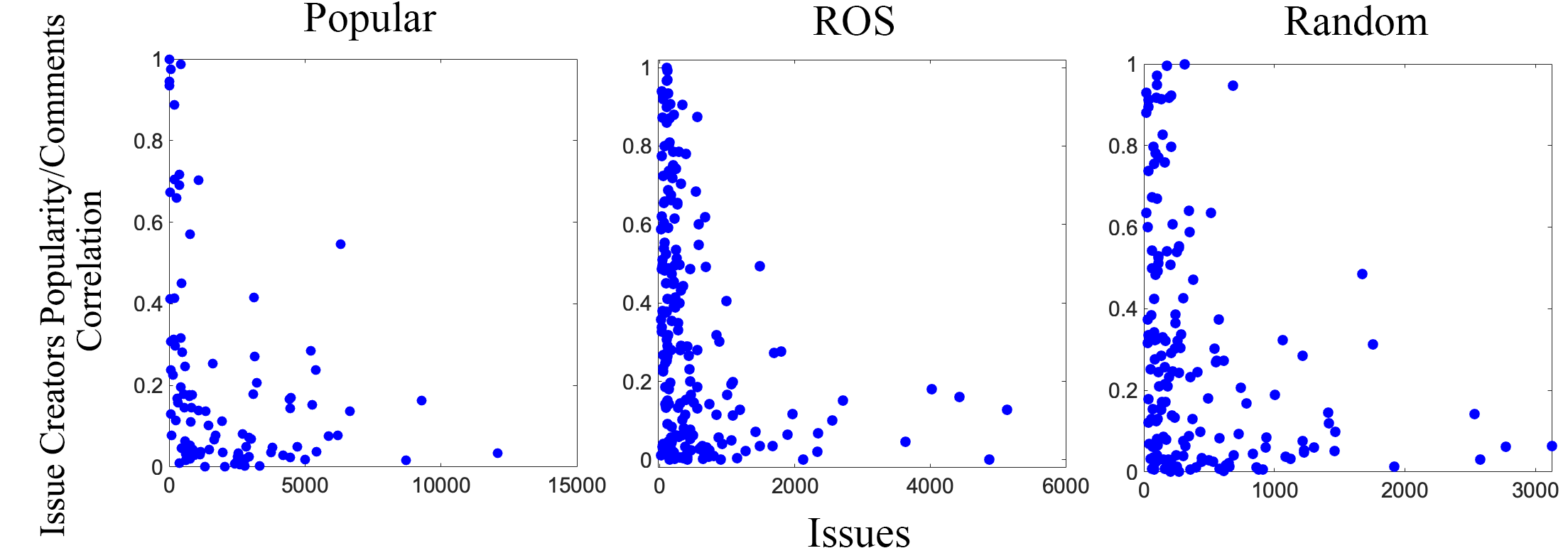}
		\caption{Correlation between Issue opener popularity and the comments they receive, sorted by Issue count.}	\label{popularity-comments}
	\end{center}
\end{figure}

Figure \ref{popularity-comments} demonstrates the relationship between the popularity of Issue openers and the number of comments that an Issue receives, sorted by Issue number. Each data point on the graph represents data from a single repository, with the higher the data point, the more comments the popular Issue openers will receive.

There is a considerable correlation between Issue openers' popularity and the number of comments they receive, as shown in Figure  \ref{popularity-comments} for repositories with fewer Issues. This means that even if a team member who opens an Issue is popular, he or she will still receive a lot of feedback. This trend can be found in most of the repositories that we have looked at.

On the other hand, we see a different pattern on the repositories with a high number of Issues. On such repositories, the popular users who open Issues, regardless of repository category, receive relatively fewer comments. In such repositories, the less popular users may receive more attention when their code is reviewed, suggesting that others may not initially approve their code.

As a result, being a popular developer in most repositories does not always imply receiving fewer corrections or getting code reviewed with no comments. This behavior indicates that most developers, including the experienced ones, will have a thorough quality check on their code.

\subsection{Threats to Validity} 
Our study assumes that using social coding platforms to enforce good software engineering practices such as review regularity, increasing community involvement in code review, and recruiting more senior software engineers to participate in the review process is likely to yield higher quality code; however, following best practices is not a guarantee of ultimate code quality.   Moreover, reminder mechanisms such as the New Issue Notifier may overload developers who are simultaneously involved in multiple projects. Team communications that occurred via Slack or videoconference were not included in our analysis.
 

\section{Conclusion}
This paper presents an analysis of the code review process when conducted using the GitHub Issue tracking mechanism.  We collected data from three types of GitHub repositories: 1) popular repositories (owned by very big tech projects, organizations, or companies) 2) ROS-related repositories (robotic specific implementations with an active user community) and 3) randomly selected public repositories.  First, an analysis was conducted of how Issue frequency and timing is affected by repository features such as age and number of contributors.  Based on these findings, we propose a mechanism, New Issues Notifier (NIN) which nudges developers towards greater regularity in the review process by prompting them to open issues during dormant periods.  This approach was simulated and executed with three threads considering different user acceptance probability on each thread; we show that even low user acceptance yields greater regularity during code review.

This paper also introduces a new metric (Issue Community Score) for evaluating community involvement and collaboration in the code review process.  Although we believe enforcing regularity is an important aspect of the code review process, it does not necessarily yield greater community involvement since it is possible for repositories with high regularity to have a low ICS score.  Fortunately even with the current GitHub Issue tracker, it appears that most repositories are successful at recruiting senior software engineers to participate in code reviews. Based on our research, we believe that it may also be beneficial to offer other recognition-based incentives to ensure a well functioning code review process with high levels of collaboration.   

\bibliographystyle{IEEEtran}
\bibliography{main}


\end{document}